\documentclass[prd, twocolumn, superscriptaddress, floatfix, nofootinbib, preprintnumbers, longbibliography]{revtex4-1}
\pdfoutput=1
\usepackage[utf8]{inputenc}
\usepackage{hyperref}
\hypersetup{colorlinks=true,citecolor=blue}
\usepackage{amssymb}
\usepackage{amsmath}
\usepackage{graphicx}
\usepackage{footmisc}
\usepackage{url}
\usepackage{multirow}
\usepackage{makecell}
\usepackage{enumerate}
\usepackage{hhline}
\usepackage{comment}
\usepackage{nicefrac}
\usepackage{physics}
\usepackage{siunitx}
\usepackage{xcolor}
\usepackage{placeins}

\def\beq{\begin{equation}}
\def\eeq{\end{equation}}
\newcommand{\bea}{\begin{eqnarray}\begin{aligned}}
\newcommand{\eea}{\end{aligned}\end{eqnarray}}

\newcommand{\oja}{\textsc{OmniJet}-$\alpha${ }}

\begin{document}

\title{Aspen Open Jets: Unlocking LHC Data for Foundation Models in Particle Physics}

\author{Oz Amram}
\email{oamram@fnal.gov}
\affiliation{Fermi National Accelerator Laboratory, Batavia, IL 60510, USA}

\author{Luca Anzalone}
\email{luca.anzalone2@unibo.it}
\affiliation{Department of Physics and Astronomy (DIFA), University of Bologna, 40127 Bologna, Italy}
\affiliation{Istituto Nazionale di Fisica Nucleare (INFN), 40127 Bologna, Italy}

\author{Joschka Birk}
\email{joschka.birk@uni-hamburg.de}
\affiliation{Institut f\"{u}r Experimentalphysik, Universit\"{a}t Hamburg, 22761 Hamburg, Germany}

\author{Darius A.\ Faroughy}
\email{darius.faroughy@rutgers.edu}
\affiliation{NHETC, Dept.\ of Physics and Astronomy, Rutgers University, Piscataway, NJ 08854, USA}

\author{Anna Hallin}
\email{anna.hallin@uni-hamburg.de}
\affiliation{Institut f\"{u}r Experimentalphysik, Universit\"{a}t Hamburg, 22761 Hamburg, Germany}

\author{Gregor Kasieczka}
\email{gregor.kasieczka@uni-hamburg.de}
\affiliation{Institut f\"{u}r Experimentalphysik, Universit\"{a}t Hamburg, 22761 Hamburg, Germany}
\affiliation{Center for Data and Computing in Natural Sciences (CDCS), 22607 Hamburg, Germany}

\author{Michael Kr\"{a}mer}
\email{mkraemer@physik.rwth-aachen.de}
\affiliation{Institut f\"{u}r Theoretische Teilchenphysik und Kosmologie, RWTH Aachen University, 52074 Aachen, Germany}

\author{Ian Pang}
\email{ian.pang@physics.rutgers.edu}
\affiliation{NHETC, Dept.\ of Physics and Astronomy, Rutgers University, Piscataway, NJ 08854, USA}

\author{Humberto Reyes-Gonzalez}
\email{humberto.reyes@rwth-aachen.de}
\affiliation{Institut f\"{u}r Theoretische Teilchenphysik und Kosmologie, RWTH Aachen University, 52074 Aachen, Germany}

\author{David Shih}
\email{shih@physics.rutgers.edu}
\affiliation{NHETC, Dept.\ of Physics and Astronomy, Rutgers University, Piscataway, NJ 08854, USA}

\begin{abstract}

Foundation models are deep learning models pre-trained on large amounts of data which are capable of generalizing to multiple datasets and/or downstream tasks. This work demonstrates how data collected by the CMS experiment at the Large Hadron Collider can be useful in pre-training foundation models for HEP.
Specifically, we introduce the \textsc{AspenOpenJets} dataset, consisting of approximately 178M high $p_T$ jets derived from CMS 2016 Open Data. We show how pre-training
the \oja\ foundation model on \textsc{AspenOpenJets} improves performance on generative tasks with significant domain shift: generating boosted top and QCD jets from the simulated JetClass dataset. 
In addition to demonstrating 
the power of pre-training of a jet-based foundation model on actual proton-proton collision data, we provide the ML-ready derived \textsc{AspenOpenJets} dataset for further public use.

\end{abstract}

\maketitle

\section{Introduction} 
While particle physics has long used machine learning techniques and is leading the way in adopting modern deep learning methods, the development and application of powerful foundation models -- which have transformed natural language processing and computer vision -- is still in its early stages. For the purposes of this work, a foundation model will be a model that has been pre-trained on a large amount of data for a specific task, and that can be used for different tasks downstream~\cite{bommasani2022opportunities}. The pre-training thus serves as a foundation, upon which the downstream tasks can be built. Well-known foundation models include large language models (LLMs) such as GPT-3~\cite{brown2020language}, BERT~\cite{devlin2019bert} and LLaMA~\cite{touvron2023llama}, image models such as Imagen~\cite{saharia2022photorealistic}, and mixed-modality models such as CLIP~\cite{radford2021clip} and ALIGN~\cite{DBLP:journals/corr/abs-2102-05918}.

There are several advantages to the use of foundation models. One example is the potential to boost the performance on small datasets. Having access to a model pre-trained on a larger, similar dataset provides a `head start' for the downstream task, allowing the model to focus on the intricacies of the smaller dataset rather than having to rediscover the general structure of the data. In addition, a foundation model can save both human and computational resources. While pre-training may be a resource-intensive task, the downstream models would require less training, less data, and less time spent on optimization compared to what would be required if these models were built from scratch. Another interesting aspect is that a trained foundation model could enable implicit sharing of data and resources between different parts of the research community, a sharing that would not otherwise be possible.

Foundation models are poised to play a central role in particle physics, especially in the context of experiments at the Large Hadron Collider (LHC). They are able to process the huge amounts of complex and diverse data generated by the LHC and have the potential to integrate multiple modalities - such as images, high-level event features or particle four-vectors - allowing more comprehensive data analyses. Current work on foundation models for particle physics has mainly focused on different event generation and classification tasks~\cite{Kishimoto:2023cys,Qu:2022mxj,Heinrich:2024sbg,Birk:2024knn,Harris:2024sra,Mikuni:2024qsr}, investigating pre-training strategies, encoding schemes and different architectures, and quantifying the transfer learning and generalization capabilities of these models. 

In this paper, we explore the potential of using large quantities of Open Data from the LHC for pre-training foundation models. The LHC experiments have made significant amounts of data publicly available~\cite{CMS:2011A, CMS:2012A, CMS:2016G,CMS:2016H, ATLAS:OPEN}. Foundation models can leverage the LHC open data for training and fine-tuning, potentially integrating different data modalities and data from different experiments. Furthermore, training foundation models on real data will result in a more accurate representation of particle physics processes than training on simulations. Finally, foundation models trained on a large variety of actual LHC data will help to promote transparency and collaboration in particle physics research. 

Here we present the \textsc{AspenOpenJets} (AOJ) dataset and study the transfer learning of foundation models obtained with this dataset.  The AOJ dataset has been constructed from the CMS 2016 `JetHT' open data~\cite{CMS:2016G,CMS:2016H} and consists of $178\,\rm M$ jets, making it the largest ML-ready dataset for LHC applications. Jets are important objects at the LHC, consisting of collimated sprays of particles. Previous open data studies of CMS jets have explored the substructure and energy flow of the jets \cite{Larkoski:2017bvj,Tripathee:2017ybi,Komiske:2019jim}  and the usefulness of open data for quark/gluon tagging \cite{Dolan:2023abg}, but this is the first study to prepare and make public a form of the CMS open data that allows easy use of the dataset for phenomenological studies, especially for training machine learning models.

As an initial proof-of-concept study, we pre-train the \oja~\cite{Birk:2024knn} foundation model on the AOJ dataset, which is expected to mostly contain QCD jets, and then fine-tune it to generate jets of different types from the JetClass~\cite{JetClass} dataset of simulated jets. We do two such fine-tunings: first for JetClass QCD jets, which occupy a different kinematic region compared to the jets in AOJ, and then for top jets, which differ significantly in several features. We show that pre-training on AOJ indeed benefits the downstream task, making it easier for the model to generate these types of jets with fewer training examples compared to training from scratch.  

The utility of CMS open data for pre-training has been previously explored in \cite{Kishimoto:2023cys}, using 1M single-lepton events at the object level which were collected by CMS in 2015. Here we enlarge the scope and complexity by several orders of magnitude to encompass 178M jet events at the full constituent level, that is, including not only the high-level features of the jets themselves, but also the features of all the particles inside the jet. Furthermore, the authors of \cite{Kishimoto:2023cys} used event selections to shape the open data into a similar topology to the intended target simulated data. In our work, no such shaping is needed. 

This paper is organized as follows. Section~\ref{sec:data} describes the CMS open data and the method used to derive AOJ. It also includes a brief description of JetClass, which will be used for the downstream task. Our application example is described in section~\ref{sec:experiment}, detailing the method and results. Finally, we present our conclusions in section~\ref{sec:conclusions}. Links to the codebase and the AOJ dataset itself can be found after the conclusions.

\section{Data}
\label{sec:data}

\begin{figure*}
    \centering
    \includegraphics[width=1\textwidth]{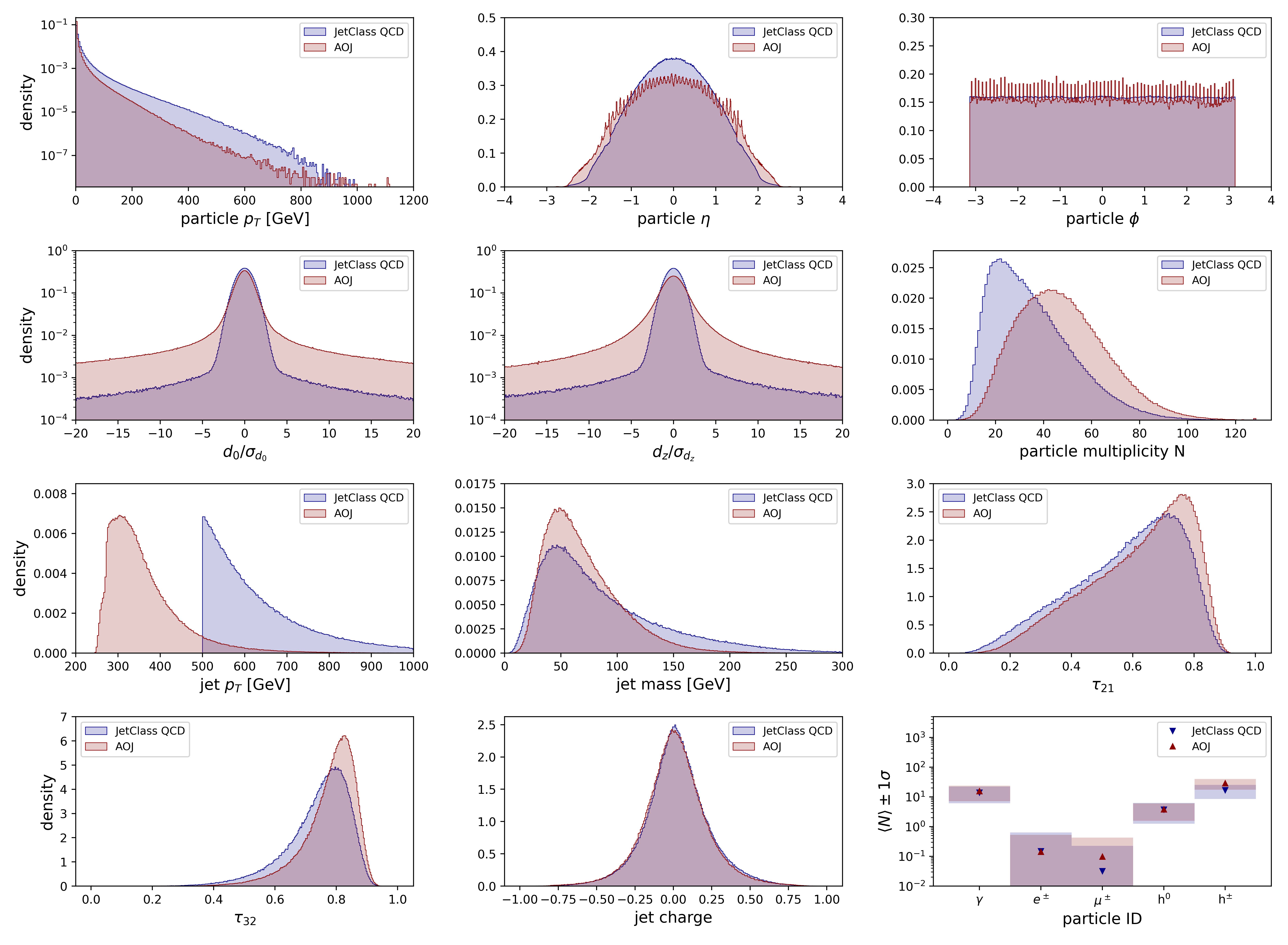}
    \caption{
        Comparison of \textsc{AspenOpenJets} (AOJ) and QCD JetClass distributions for particle and jet level observables. Here $d_0$ and $d_z$ represent the transverse and longitudinal impact parameter respectively, with $\sigma_{d_0}$ and $\sigma_{d_z}$ being their uncertainties, and $\tau_{21}$ and $\tau_{32}$ are N-subjettiness variables. The last panel shows the average number of jet constituents for each particle species, accompanied by the corresponding $1\sigma$ standard deviation bands.
        }
    \label{fig:constituent_aoj}
\end{figure*}

\subsection{CMS open data}
\label{sec:cmsopen}
The CMS experiment \cite{CMS:2008xjf} is a large general-purpose detector at the LHC. The CMS collaboration has recently released 16.4 fb$^{-1}$ of data from their 2016 runs (G and H) \cite{CMS:2016G,CMS:2016H} and a corresponding set of simulations. In contrast to previous data releases \cite{CMS:2010B,CMS:2011A,CMS:2012A}, this data has been provided in the simplified \texttt{MINIAOD} format~\cite{Petrucciani:2015gjw}, as well as the even further simplified \texttt{NANOAOD} format~\cite{Peruzzi:2020}. In these simplified analysis formats, the central CMS reconstruction steps have been applied, discarding much of the low-level content of the event in favor of a simplified description of the high-level objects (jets, electrons, muons, etc.) commonly used in data analysis. These formats are therefore significantly easier to use for those outside of the collaboration.

The inner layers of CMS consist of a silicon pixel and silicon strip tracker used to measure charged particle trajectories.
These are followed by a lead tungstate crystal electromagnetic calorimeter and a brass and scintillator hadronic calorimeter. These detector elements are enclosed in a large solenoid, which produces a magnetic field of 3.8\,\unit{T}.
Muon stations are located outside the iron return yoke of the magnet. Events of interest are selected using a two-stage trigger system. The first stage (L1) is implemented in custom hardware to select potentially interesting events within a fixed latency of about 4\,$\mu$s~\cite{CMS:2020cmk} using information from the calorimeters and muon stations. Events accepted by the L1 system are sent to the high-level trigger (HLT), which consists of a dedicated computing cluster which runs a speed-optimized version of the full reconstruction. The HLT uses this more accurate reconstruction of the event to select events to be saved at a rate of approximately 1\,\unit{kHz}~\cite{CMS:2016ngn}. 

CMS uses a particle-flow algorithm~\cite{CMS:2017yfk}, which combines the information from the different detector systems to reconstruct and identify each individual particle in an event. Particles are classified as either muons, electrons, photons, charged hadrons or neutral hadrons.  Jets are clustered from the particle-flow candidates in an event using the anti-$k_t$ jet finding algorithm~\cite{Cacciari:2008gp,Cacciari:2011ma}. The typical large-radius jet size used in CMS for jet substructure searches and measurements is $R=0.8$, commonly referred to as AK8 jets. For AK8 jets the pileup-per-particle identification algorithm (PUPPI)~\cite{Sirunyan:2020foa,Bertolini:2014bba} is used to mitigate the effect of pileup. The PUPPI algorithm assigns a weight to each particle based on the probability that it originates from the primary vertex or from pileup.  The momentum of AK8 jets is determined from the weighted vector sum of all particles inside the jet. Additional corrections are applied based on simulation and data-driven measurements to better calibrate the measured jet momentum to the true momentum of the particle from the hard process~\cite{CMS:2016lmd}. 

\subsection{Extracting Aspen Open Jets}
\label{sec:aojextract}
The \texttt{NANOAOD} format does not natively store information of the jet constituents. However, an extended version (\texttt{PFNANO}), which does include this information, can be easily produced from \texttt{MINIAOD} using tools provided by CMS \cite{PFNANO}.
We therefore, for all data and simulations samples used, start with the \texttt{MINIAOD} format, process it to \texttt{PFNANO} and then apply our selections to select the jets of interest for our dataset.

In order to construct the largest sample of jets possible, we consider all events from the `JetHT'~\cite{CMS:2016G,CMS:2016H} Run-2 dataset released by CMS to this date.
The `JetHT' dataset includes all events that have been selected by one of the HLT triggers related to jet momenta or total event hadronic activity (HT). Events are required to either have a high scalar sum of jet transverse momenta, or contain at least one or two jets with high energy. For the exact details of all different trigger paths used, see Refs.~\cite{CMS:2016G,CMS:2016H}. Because we do not specify a specific trigger, the sample consists of events from an inclusive `or' of any of the available triggers of this dataset. We note that this choice works well for building a large dataset for the purposes of foundation model training, but may introduce subtle kinematic biases resulting from the combined thresholds of the different triggers, which would complicate its use in a physics-based analysis. 

Events are required to pass several data-quality filters commonly used within CMS to remove events that are likely to cause problems in the reconstruction. AK8 jets are required to have transverse momentum greater than 300\,\unit{GeV}\footnote{The jet transverse momentum ($p_T$) distribution in fig.~\ref{fig:constituent_aoj} extends down to ~250 \unit\,{GeV} as it is computed directly from vector sum of the jet constituents. Whereas, 300\,\unit{GeV} threshold is applied to the jet after additional corrections have been applied to the jet energy.}.
The jet $p_T$ spectrum is steeply falling, and the 300\,\unit{GeV} threshold was chosen to balance having a large dataset of jets for the pre-training task with the consideration of matching the high-momentum kinematic regime typical of typical jet substructure studies. 
Jets are also required to have an absolute value of pseudorapidity ($\eta$) less than 2.5, so that they fall within the range of the CMS tracker. The AK8  jets are also required to pass standard CMS quality criteria used to reject leptons reconstructed as jets and jets that may be poorly reconstructed. We construct our sample out of all AK8 jets that pass these criteria, with no limitation on the number of selected jets per event. 

Applying these selections to the CMS open data yields a dataset of 178 million jets, which we call the \textsc{AspenOpenJets} (AOJ). Due to the large cross section of QCD processes, the vast majority of these jets are expected to originate from light quarks and gluons. 
A small fraction of the jets are expected to originate from the decays of boosted heavy resonances. 
A rough estimate of the number of boosted fully merged jets produced by W, Z, top, and Higgs bosons in the $b\bar{b}$ decay mode are estimated with Monte Carlo (MC) simulations provided by CMS \cite{CMS:2016MC}, with the remainder assumed to originate from QCD. The numbers are reported in Table~\ref{tab:num_jet_types}. More details on the MC samples used are given in Appendix~\ref{app:mc}.

\begin{table}[]
    \centering
    \begin{tabular}{c|c}
        Jet type & Approx. Number \\
        \hline
         QCD & $1.8\times 10^{8}$\\
         $W$   & $6.3\times 10^{5}$ \\
         $Z$   & $1.8\times 10^{5}$ \\
         top & $1.3\times 10^{5}$ \\
         $H \to b\bar{b}$ & 500 \\
    \end{tabular}
    \caption{Breakdown of the expected number of jet types in the AOJ dataset from MC simulations.
    }
    \label{tab:num_jet_types}
\end{table}

For each jet we store its transverse momentum ($p_T$), pseudorapidity ($\eta$), and azimuthal angular coordinate ($\phi$). 
We also store its mass, groomed with the softdrop algorithm \cite{softdrop} as computed within the CMS reconstruction.
Up to 150 constituents of the jet are stored. For each constituent, its 4-momentum is stored in the format $(p_x, p_y, p_z, E)$. 
We additionally store its transverse impact parameter ($d_0$) and longitudinal impact parameter ($d_z$) with their uncertainties, the charge of the candidate, its particle-ID (PID) in the PDG format\footnote{Note that neutral hadrons $({\rm h}^0)$ are assigned the ${\rm PID}=130$ of the neutral kaon $K_L^0$ while positively/negatively charged hadrons $({\rm h}^\pm)$ are assigned ${\rm PID}=\pm211$ of the charged pion.} \cite{ParticleDataGroup:2024cfk} and its weight from the PUPPI algorithm. We also include additional jet substructure quantities computed within the CMS reconstruction, including the number of constituents in the jet, $N$-subjettiness variables \cite{nsubjettiness}, various jet-tagging observables from the CMS implementation of ParticleNet \cite{particlenet, particlenet-cms} and a regression of the jet mass from ParticleNet \cite{particlenet-cms}.

\begin{figure}
    \centering
    \includegraphics[width=0.75\linewidth]{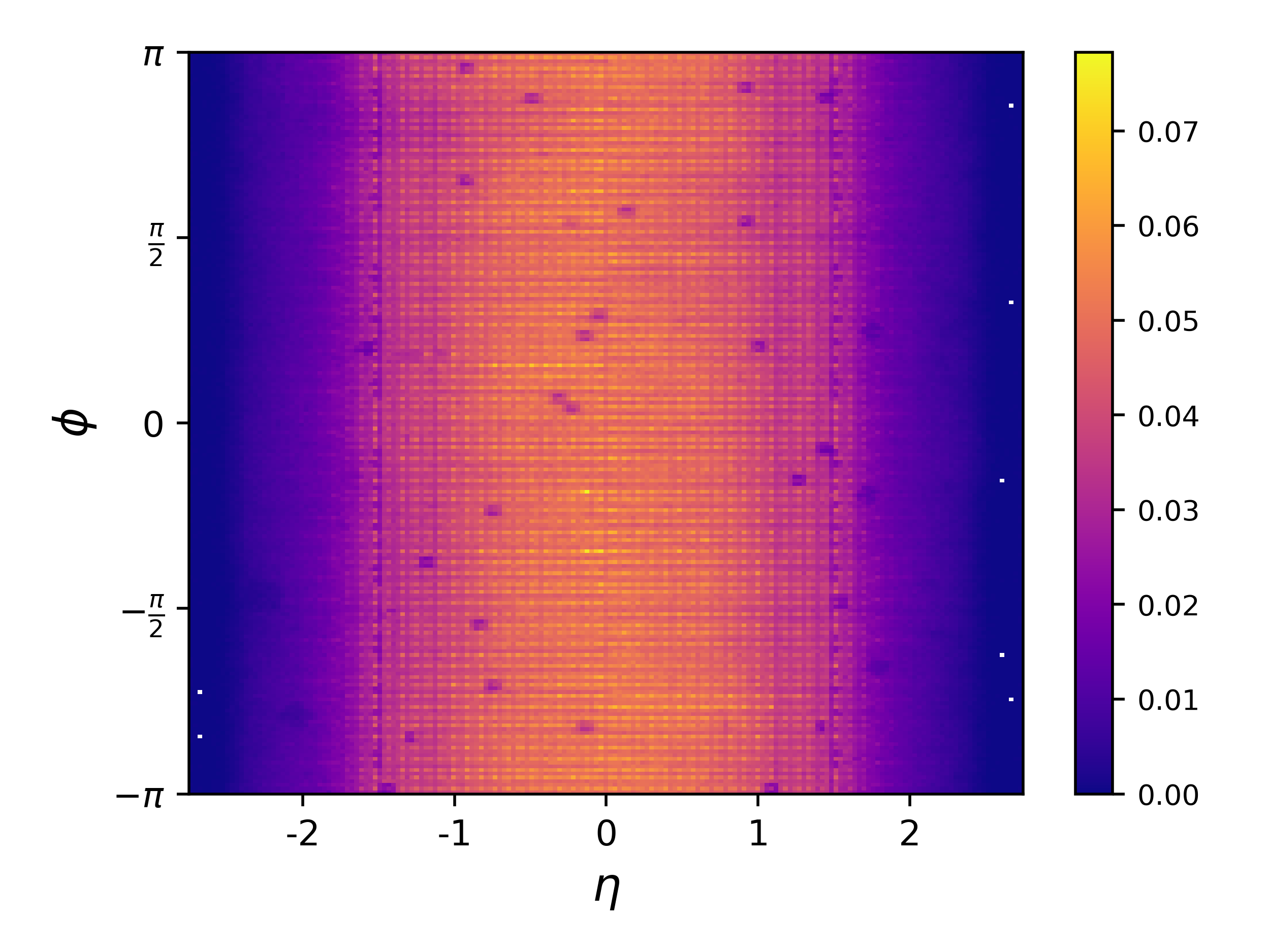}
    \caption{{The particle-level positional distribution of the \textsc{AspenOpenJets} in the $\eta-\phi$ plane, illustrating the granularity of the CMS detector. Note the appearance of vertical features indicating the endcaps at $|\eta| \sim 1.6$ and multiple ``dead cells'' throughout the detector.}
    }
    \label{fig:aoj_eta_phi}
\end{figure}

In fig.~\ref{fig:constituent_aoj} we show the distributions (red histograms) of several particle and jet-level features, based on a sub-sample of $700\,\rm K$ jets. Interestingly, the effects of detector granularity are seen as spikes in the particle $\eta$ and $\phi$ distributions. In fig.~\ref{fig:aoj_eta_phi} we plot a 2D histogram in the $\eta-\phi$ plane. A closer look reveals a grid-like structure corresponding to the granularity of the detector. Furthermore, there are vertical features indicating the endcaps at $\abs{\eta} \sim 1.6$ and several ``dead cells" throughout the detector. The distribution of the subjettiness ratio $\tau_{32}$ is consistent with the expectation that the AOJ is predominantly composed of QCD jets. The last panel in fig.~\ref{fig:constituent_aoj} shows the average number of constituents per jet for each PID, accompanied by the corresponding 1$\sigma$ standard deviation bands. 

\subsection{JetClass}
\label{sec:jetclass}
The JetClass~\cite{JetClass} dataset is based on simulations and was first introduced in~\cite{Qu:2022mxj}. It contains both particle-level and jet-level information for $125\,\rm M$ simulated jets, distributed over 10 different jet types and divided into a training~($100\,\rm M$), validation~($5\,\rm M$) and test~($20\,\rm M$) set. The simulation is performed with \textsc{MadGraph5\_aMC@NLO}~\cite{Alwall:2014hca} for the hard process, followed by \textsc{Pythia}~\cite{Sjostrand:2014zea} for parton showering and hadronization, and \textsc{Delphes}~\cite{deFavereau:2013fsa} with the CMS detector~\cite{CMS:2008xjf} card for the detector response. Jets are required to have a transverse momentum of $500\,\mathrm{GeV} < p_T < 1000\,\mathrm{GeV}$ and a pseudorapidity $\eta<2$, which is more restrictive than the criteria used for AOJ. We overlay the JetClass QCD particle- and jet-level distributions (blue histograms) in fig.~\ref{fig:constituent_aoj} to facilitate easy comparison with the corresponding AOJ distributions.

\section{Experiment}
\label{sec:experiment}
In this section, we demonstrate the usefulness of the AOJ dataset in ML-related applications. In particular, we pre-train a foundation model on the full AOJ dataset.

\subsection{Method}
\label{sec:method}

The foundation model chosen for this work is \oja~\cite{Birk:2024knn}, a GPT-style~\cite{Radford2018ImprovingLU} model that predicts the next token in a sequence. The features of the jet constituents\footnote{The jet constituents are $p_T$ ordered and up to the first 128 constituents are considered when training the models.} are first converted into integer tokens using a VQ-VAE~\cite{oord2018neural,bao2022beit,Heinrich:2024sbg,huh2023straightening} model, representing jets as sequences of tokens. The backbone of \oja is trained to predict the next token in these sequences. After pre-training, the model can generate new sequences autoregressively, which are then decoded by the frozen VQ-VAE back into physical space. Additionally, the pre-trained backbone weights can be loaded and used for other tasks, such as jet classification. In this work, we focus on dataset transfer learning for jet generation and leave the exploration of task switching for future studies.

The features used for training are the constituent kinematics ($p_T$, $\Delta\eta$, $\Delta\phi$), where $\Delta\eta = \eta - \eta_{\text{jet}}$ and $\Delta\phi = \phi - \phi_{\text{jet}}$ represent the relative coordinates with respect to the jet axis. Before tokenization, these features were preprocessed using standardization\footnote{Standardization is a common technique used to bring feature values into a manageable range for neural networks. By subtracting the mean and dividing by the standard deviation, the data is transformed to having a mean of zero and standard deviation of one.}. The AOJ dataset was tokenized using a codebook size of 8192 tokens, employing VQ-VAE hyperparameters as outlined in \cite{Birk:2024knn}, with two modifications: the number of attention heads was reduced from 8 to 4 without sacrificing performance, and the AdamW optimizer was replaced with the Ranger optimizer \cite{Ranger}. A larger codebook with $\sim\!32\,\rm K$ tokens was also tested but produced results comparable to the $8\,\rm K$ configuration. Therefore, we only present results for the $8\,\rm K$ tokenization. The backbone model was trained on the full $178\,\rm M$ AOJ dataset with hyperparameters as specified in \cite{Birk:2024knn}. The model was trained for 900K steps on 12 GPUs with a batch size of 256 per GPU, and the final state at the end of the training  was selected as the backbone model.

To evaluate the transfer learning capabilities of \oja, we devised a downstream task to test its ability to generate jets of a different type than those it was pre-trained on, and to adapt to datasets with different kinematic cuts. The fine-tuning process involved loading the pre-trained backbone weights and retraining it using jets from the JetClass~\cite{JetClass} dataset. To systematically evaluate the performance of the model, we trained it on six datasets with $10^2$, $10^3$, $10^4$, $10^5$, $10^6$, $10^7$ jets to study the effect of dataset size $D$, and fine-tuned the model separately on QCD jets ($q/g$) and top-quark jets ($t \to bqq'$) to evaluate its adaptability to different jet substructure types. For each configuration, the model with the lowest validation loss during fine-tuning was selected for jet generation. To provide a baseline for comparison, we used a separate VQ-VAE tokenizer and trained generative models with the same backbone architecture from scratch using randomly initialized weights. This VQ-VAE is the one from Ref.~\cite{Birk:2024knn} that was trained on all 10 classes of the JetClass dataset. The generative models were trained on the tokenized QCD and top-quark samples. This approach allowed us to quantify the performance improvement achieved by pre-training on the AOJ dataset.

Both training strategies used a batch size of 256\footnote{For training samples of 100 jets, the batch size was reduced to 100.} per GPU\footnote{We used a single GPU to train the models with training sample sizes less than or equal to $100\,\rm K$, and multiple GPUs for models trained on more data.} and a patience of 10 epochs. For training samples with sizes up to $100\,\rm K$ jets, the validation set was the same size as the training set. For larger samples, we used validation set sizes of $200\,\rm K$  jets for $1\,\rm M$ training samples and $500\,\rm K$  jets for $10\,\rm M$  training samples.

\subsection{Performance metrics}

We consider two complementary evaluation metrics that compare the 1D distribution of a high-level observable $\mathcal{O}$ between the generated jets and the target JetClass reference data:
\begin{itemize}
    \item Kullback-Leibler divergence:
    \begin{equation}
    {\rm KL}^\mathcal{O}(P||Q) = \sum_x P(x) \log\left(\frac{P(x)}{Q(x)}\right)\,.    
    \end{equation}
    The probability densities $P(x)$ and $Q(x)$ for the observable $\mathcal{O}$ are obtained from the normalized histogram bin counts and ${\rm KL}^\mathcal{O}(P||Q)$ is computed using \texttt{scipy.stats.entropy}~\cite{2020SciPy-NMeth}. Note that ${\rm KL}^\mathcal{O}(P||Q)$ is antisymmetric. In our results, we take $P$ to be the generated distribution and $Q$ to be the reference JetClass distribution.
    
    \item Wasserstein-1 distance: 
    \begin{equation}
    W^\mathcal{O}_1(P,Q) = \min_{\pi \in \Pi}\,\sum_{x,y} \abs{x-y} \ \pi(x,y)\,.      
    \end{equation}
    Here $\Pi(P,Q)$ is the space of all joint `coupling' distributions whose marginals are the observable $\mathcal{O}$ distributions $P$ and $Q$. This quantity is computed with \texttt{scipy.stats.wasserstein\_distance}~\cite{2020SciPy-NMeth}. Unlike the ${\rm KL}^\mathcal{O}(P||Q)$ divergence, $W^\mathcal{O}_1(P,Q)$ is a symmetric similarity metric based on optimal transport, i.e. the optimal cost plan that morphs $P$ into $Q$.
\end{itemize}

While further detailed performance evaluations such as classifier tests~\cite{Krause:2021ilc, Kansal:2022spb, Das:2023ktd} could also be envisioned, this paper is first and foremost demonstrating a use case for the AOJ dataset in the context of foundation models. As such, we are rather concerned with the overall qualitative behavior, for which the 1D distributions suffice. 

\subsection{Results} 
\label{sec:results}

\begin{figure*}
    \centering
    \includegraphics[width=1\linewidth]{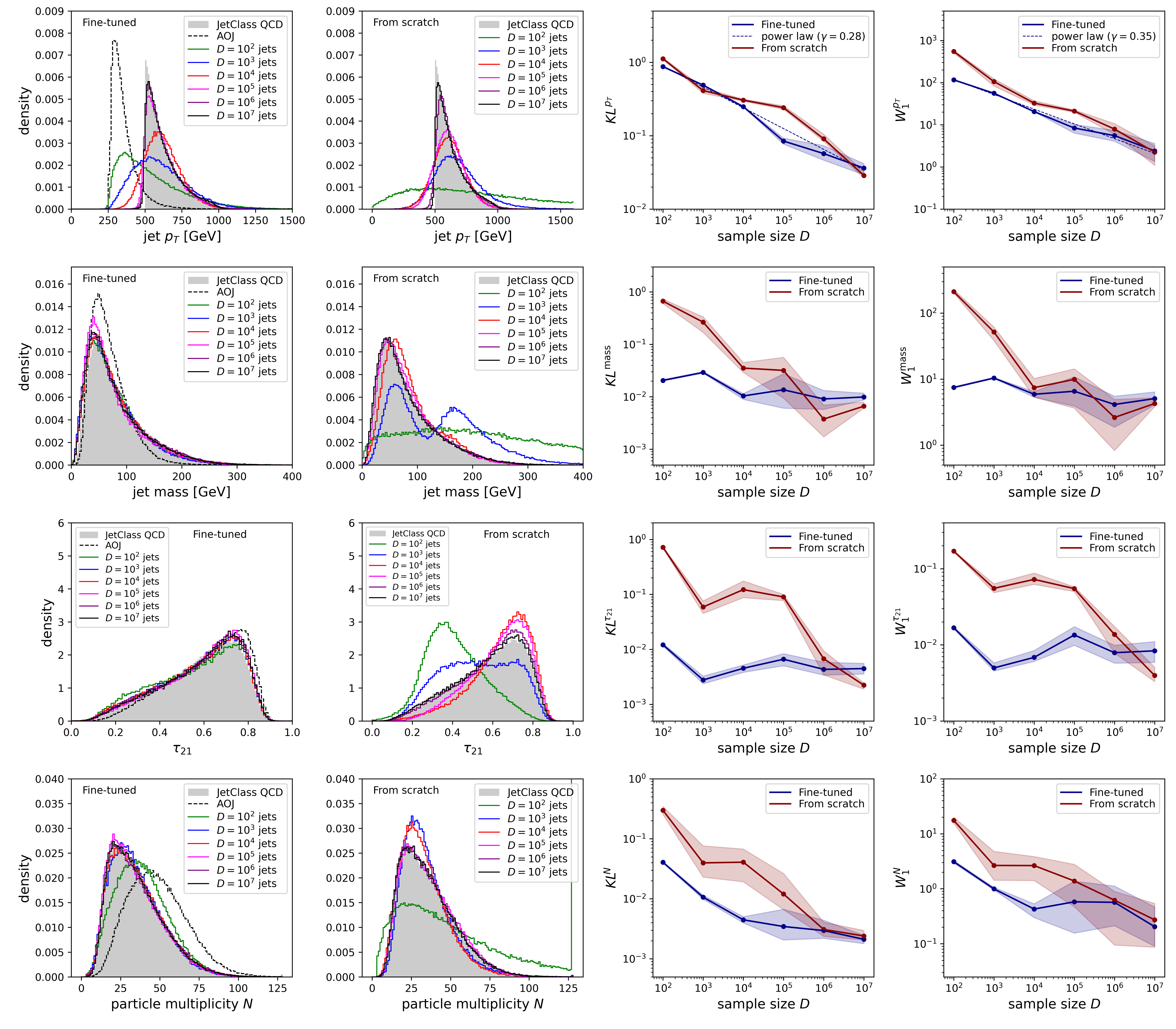}
    \caption{
        A comparison of the generation quality of models trained on QCD jets
        from JetClass, for different training sample sizes $D$. The performance
        of a foundation model pre-trained on AOJ (far left) is compared to a
        model trained from scratch (center left). The generation quality is
        compared across several high level features of the jets. Two
        quantitative metrics, the Kullback-Leibler divergence (center right) and
        Wasserstein-1 distance (far right), are computed as a function of the
        training sample size $D$ to compare how well the generated jets from
        each model matches the target distribution from JetClass. We also report the
    mean value and the envelopes over 5 trainings with different random
    seeds. When applicable, we also show power-law fits $\propto D^{-\gamma}$ to the metrics.
    }
    \label{fig:gen_finetune_vs_fromscratch_qcd}
\end{figure*}

\begin{figure*}
    \centering
    \includegraphics[width=1\linewidth]{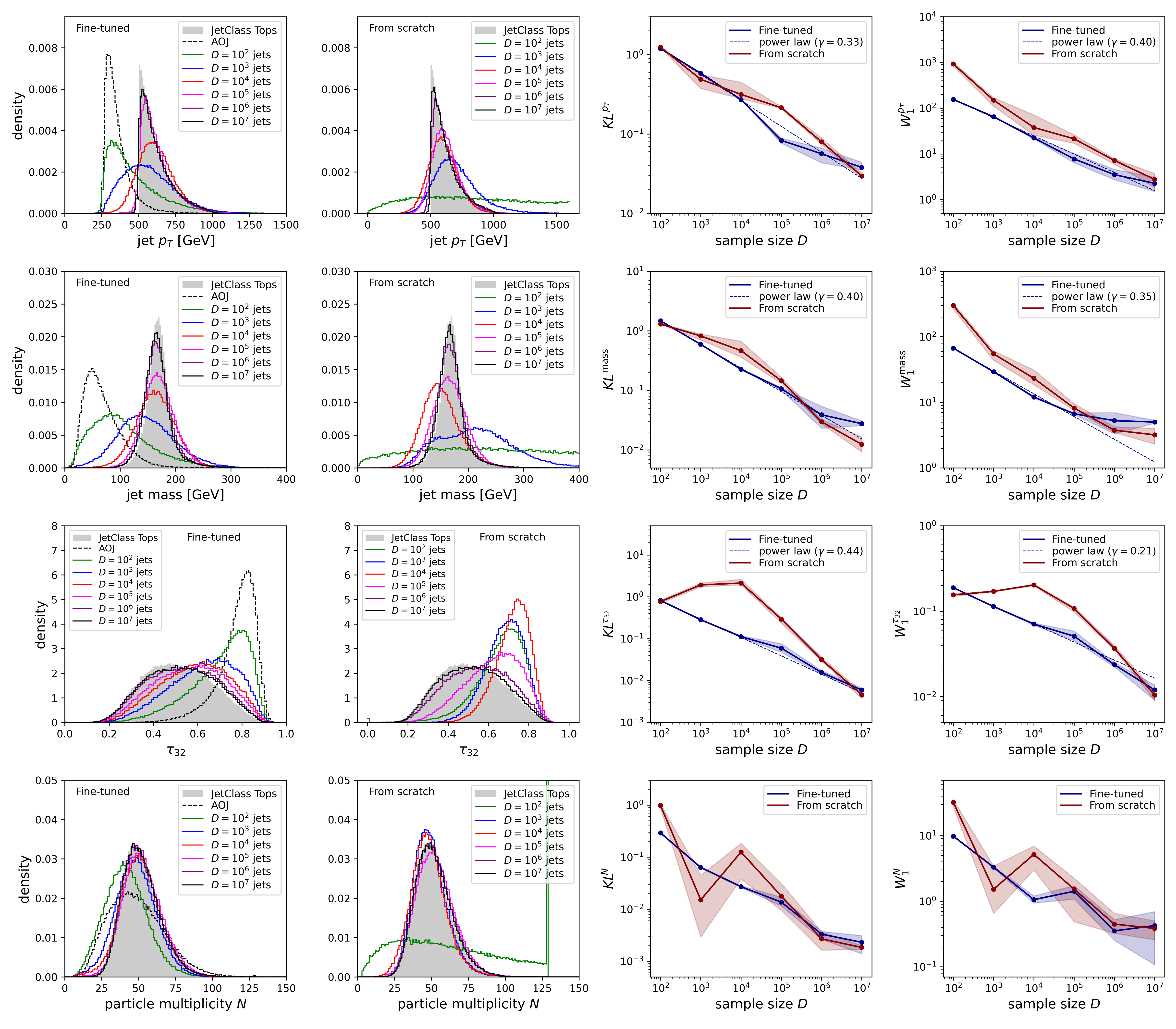}
    \caption{
A comparison of the generation quality of models trained on top-quark jets from JetClass, for different training sample sizes $D$. The performance of a foundation model pre-trained on AOJ (far left) is compared to a model trained from scratch (center left). The generation quality is compared across several high level features of the jets. Two quantitative metrics, the Kullback-Leibler divergence (center right) and Wasserstein-1 distance (far right), are computed as a function of the training sample size $D$ to compare how well the generated jets from each model matches the target distribution from JetClass. We also report the
    mean value and the envelopes over 5 trainings with different random
    seeds. When applicable, we also show power-law fits $\propto D^{-\gamma}$ to the metrics.}\label{fig:gen_finetune_vs_fromscratch_tops}
\end{figure*}

A comparison of the two training strategies --``fine-tuned" and ``from scratch"-- is presented in fig.~\ref{fig:gen_finetune_vs_fromscratch_qcd} for QCD and in fig.~\ref{fig:gen_finetune_vs_fromscratch_tops} for top-quark jets. The rows show various high-level distributions, including jet $p_T$, jet mass, subjettiness ratios $\tau_{21}$ and $\tau_{32}$, and particle multiplicity $N$ (i.e.\ the number of constituents in a jet).
The left-most column in each figure displays the distributions generated by the fine-tuned models, while the second column shows the distributions generated by the models trained from scratch. Each plot also includes the reference JetClass distribution ($200\,\rm K$ reference jets) as a filled histogram and, for additional context, the AOJ distributions (dotted lines) in the left-most column.

Overall, we find that the pre-trained models are able to achieve a performance
gain when using a fraction of the total fine-tuning data. This can be seen in
the third and last columns of fig.~\ref{fig:gen_finetune_vs_fromscratch_qcd} and
fig.~\ref{fig:gen_finetune_vs_fromscratch_tops}, where the fine-tuned models
(blue lines) typically achieve better\footnote{However, there are instances
where models trained from scratch perform equally well on a single metric within uncertainties.
For example, this occurs with ${\rm KL}^{p_T}$ for training sample size
$D=10^3$. This behavior can be attributed to the definition of KL divergence,
which penalizes high-probability regions in the generated data where the true
probability distribution is low. Therefore, it is crucial to evaluate
performance using multiple metrics to gain a comprehensive understanding.}
metric scores than the models trained from scratch (red lines). This is apparent
for the KL divergence and Wasserstein-1 metrics across different training sample
sizes below $1\,\rm M$ jets.  However, as the available training data exceeds $1\,\rm M$ jets, we find that the fine-tuning begins to saturate, and the fine-tuned and from-scratch models perform similarly within uncertainties. In some features, the fine-tuned models even exhibit a slight decrease in performance compared to those trained entirely from scratch, indicating a diminishing return from pre-training at larger dataset scales. The reasons for this are unclear and would require further study. One possibility is that the AOJ tokenization is suboptimal for describing JetClass and this is exposed with enough JetClass training data. Another possibility is that the pre-training itself can harm model performance by locking in the weights to a suboptimal part of parameter space, a phenomenon that has been observed previously in the ML literature \cite{DBLP:journals/corr/abs-2102-01293} and was termed ``ossification". 

We now examine in more detail the transfer learning from AOJ to JetClass for QCD
jets, as shown in fig.~\ref{fig:gen_finetune_vs_fromscratch_qcd}. The primary
challenge in fine-tuning arises from the differences in the domains of the jet
$p_T$ distributions and particle multiplicities between the two datasets.
JetClass imposes a narrower $p_T$ window between $500$~GeV and $1000$~GeV and
tends to have around $\sim30$ particles per jet on average, while AOJ only
applies a lower cut around $250$~GeV and has around $\sim50$ particles per jet
on average. When fine-tuning, the pre-trained model must adapt to these
differences and learn to generate jets with fewer particles and the new
kinematic cuts. The results for the $p_T$ distributions and particle
multiplicities show how the fine-tuned model smoothly interpolates between the
AOJ and JetClass as a function of the training sample size $D$. Furthermore, the
generation quality improves with increasing training sample size, achieving
almost optimal results with a reduced sample size of $100\, \rm K$ jets for the
$p_T$ distribution and $10\,\rm K$ for the multiplicities. In contrast, the
randomly initialized models trained from scratch require a significantly larger
dataset, approximately an order of magnitude more jets, to achieve optimal
performance. For the jet mass and the substructure observable \(\tau_{21}\), the
fine-tuning performance saturates with a remarkably small number of training
samples, resulting in an almost flat dependence of the metrics on $D$.
Specifically, fine-tuning requires only about $1000$ jets to reach optimal
performance, whereas training from scratch requires significantly more data to
achieve comparable results. This is not too surprising, as the transfer from AOJ
to JetClass for these features does not require much effort, given the
similarity in the shapes of these distributions.

Next, we examine the results in fig.~\ref{fig:gen_finetune_vs_fromscratch_tops},
which correspond to top-quark jet generation. Since the AOJ dataset is
completely dominated by QCD jets, the transfer learning task between AOJ and
JetClass becomes more challenging due to the change in jet type. For instance,
fine-tuning must map the low-mass Sudakov peak characteristic of QCD jets to the
top-quark resonance peak at $m_t = 175$ GeV. Additionally, the model must
transition from generating 1-prong QCD jets to 3-pronged top-quark jets, as
reflected in the horizontal gap between the peaks in the $\tau_{32}$
substructure distributions. Despite these obstacles, we find that transfer
learning consistently outperforms training from scratch when using smaller
datasets. For example, the top-quark mass, represented by the mode in the jet
mass distribution, is learned with good precision by the fine-tuned model using
a training dataset of only $10\,\rm K$ jets. In contrast, the model trained from
scratch requires an order of magnitude more data to achieve comparable
performance. A similar trend is observed for the $\tau_{32}$ distribution, where
transfer learning morphs the AOJ jets into JetClass top-quark jets, achieving
competitive results with only $100\,\rm K$ training jets. In comparison, models
trained from scratch fail to generate jets with a reliable substructure for
training sample sizes below $1\,\rm M$ jets.

\subsection{Power-law scaling}

Finally, we observe that the performance metrics for the fine-tuned models, when
not in the saturated regime, seem to follow a power-law scaling of the form
$D^{-\gamma}$ with respect to the training sample size $D$. Neural scaling laws, such as these, have attracted broad interest and have been analyzed extensively in various settings in recent years, see \cite{epoch2023scalinglawsliteraturereview} for a nice overview of the ML literature and original references. In particular, they have been observed in the context of jet classification \cite{Batson:2023ohn}. We find that the exponents for a wide range of high-level observables are consistently within the range
$0.2 < \gamma < 0.45$ (see dashed lines in fig.
\ref{fig:gen_finetune_vs_fromscratch_qcd} and
\ref{fig:gen_finetune_vs_fromscratch_tops}). This scaling behavior holds for
both the KL and Wasserstein-1 metrics and applies to both QCD and top-quark,
suggesting a form of {\it universality}. The only exception to this scaling is the particle multiplicity distribution, which shows a more irregular dependence on the training sample size. 
Previous scaling law behavior has been observed as a function of the loss of the model, not on such hand-crafted metrics. 
However, due to the different tokenizers used in the from-scratch and the AOJ fine-tuned model, the losses for the generation objective are not directly comparable, making statements of true scaling comparisons difficult. 
We leave further investigation of this apparent scaling behavior to future work.

In contrast, the randomly initialized models trained from scratch are not seen to follow
such universal scaling across most physical observables. The reason behind this may be that these operate in the small-data regime \cite{hestness2017deep}, where their performance is limited by the lack of sufficient training data. In this regime, models can only perform marginally better than random guessing, as each additional sample provides limited new information for learning. The breaking of scaling is particularly evident for the jet substructure
features, such as $\tau_{21}$ for QCD and $\tau_{32}$ for tops, and to a lesser
extent for the jet mass, where the performance metrics exhibit  irregular and
dataset-dependent scaling as the training sample size increases. Only after the model has seen sufficient training data, approximately
around $1\,\rm M$ jets, does the training stabilize, resulting in reliable jet
substructure and jet mass modeling. While further studies into these scaling laws need to be performed, the observed discrepancy highlights the possible advantage of
pre-training in achieving a more robust and systematic improvement in generative
performance.
 
\section{Conclusion}
\label{sec:conclusions}

We are releasing the \textsc{AspenOpenJets} (AOJ) dataset, which comprises 178 million
jets from CMS open data, in a standard machine learning-friendly format. This
dataset provides an excellent resource for exploring foundation models and other
machine learning techniques that use large, unlabelled datasets for
pre-training.

In addition, we present the first jet-based foundation model in high-energy physics trained on real collider data from the AOJ dataset. By using this rich dataset to pre-train a foundation model, we show that fine-tuning for specific tasks - such as generating jets of different types in different kinematic regions - yields significant performance improvements over models trained from scratch, and boosts the achievable downstream performance on small labeled datasets. In particular, the generative performance of our foundation model scales predictably with training sample size, facilitating effective transfer learning. This demonstrates the power of foundation models to provide improved performance even over significant domain shifts.

We hope that this study will encourage further engagement with open data from
LHC experiments, and promote advances that bridge phenomenological studies and
deployment in actual  experiments.

\begin{table*}[t]
    \centering
    \scalebox{0.65}{
    \begin{tabular}{c c c}
        \large{\textbf{Process}} & \large{\textbf{CMS Dataset Name}} & \large{\textbf{Cross Section (pb)}} \\
        \hline
        data & /JetHT/Run2016G-UL2016\_MiniAODv2-v2/MINIAOD & - \\
             & /JetHT/Run2016H-UL2016\_MiniAODv2-v2/MINIAOD & - \\ 
        
        \hline
        QCD & /QCD\_Pt\_300to470\_TuneCP5\_13TeV\_pythia8/RunIISummer20UL16MiniAODv2-106X\_mcRun2\_asymptotic\_v17-v1/MINIAODSIM & 7823 \\
            & /QCD\_Pt\_470to600\_TuneCP5\_13TeV\_pythia8/RunIISummer20UL16MiniAODv2-106X\_mcRun2\_asymptotic\_v17-v1/MINIAODSIM & 648.2 \\
            & /QCD\_Pt\_600to800\_TuneCP5\_13TeV\_pythia8/RunIISummer20UL16MiniAODv2-106X\_mcRun2\_asymptotic\_v17-v1/MINIAODSIM & 186.9 \\
            & /QCD\_Pt\_800to1000\_TuneCP5\_13TeV\_pythia8/RunIISummer20UL16MiniAODv2-106X\_mcRun2\_asymptotic\_v17-v1/MINIAODSIM & 32.3 \\ 
            
            $\mathrm{t\bar{t}}$ & /TTToHadronic\_TuneCP5\_13TeV-powheg-pythia8/RunIISummer20UL16MiniAODv2-106X\_mcRun2\_asymptotic\_v17-v1/MINIAODSIM & 378.5 \\
             & /TTToSemiLeptonic\_TuneCP5\_13TeV-powheg-pythia8/RunIISummer20UL16MiniAODv2-106X\_mcRun2\_asymptotic\_v17-v1/MINIAODSIM & 365.2 \\
             
             W + jets & /WJetsToQQ\_HT-400to600\_TuneCP5\_13TeV-madgraphMLM-pythia8/RunIISummer20UL16MiniAODv2-106X\_mcRun2\_asymptotic\_v17-v2/MINIAODSIM & 315.6 \\
             & /WJetsToQQ\_HT-600to800\_TuneCP5\_13TeV-madgraphMLM-pythia8/RunIISummer20UL16MiniAODv2-106X\_mcRun2\_asymptotic\_v17-v2/MINIAODSIM & 68.6 \\
             & /WJetsToQQ\_HT-800toInf\_TuneCP5\_13TeV-madgraphMLM-pythia8/RunIISummer20UL16MiniAODv2-106X\_mcRun2\_asymptotic\_v17-v2/MINIAODSIM & 34.9 \\

             Z+ jets & /ZJetsToQQ\_HT-400to600\_TuneCP5\_13TeV-madgraphMLM-pythia8/RunIISummer20UL16MiniAODv2-106X\_mcRun2\_asymptotic\_v17-v2/MINIAODSIM & 145.4 \\
             & /ZJetsToQQ\_HT-600to800\_TuneCP5\_13TeV-madgraphMLM-pythia8/RunIISummer20UL16MiniAODv2-106X\_mcRun2\_asymptotic\_v17-v2/MINIAODSIM & 34.0 \\
             & /ZJetsToQQ\_HT-800toInf\_TuneCP5\_13TeV-madgraphMLM-pythia8/RunIISummer20UL16MiniAODv2-106X\_mcRun2\_asymptotic\_v17-v2/MINIAODSIM & 18.7 \\

             H$\to \mathrm{b\bar{b}}$ & /GluGluHToBB\_Pt-200ToInf\_M-125\_TuneCP5\_MINLO\_13TeV-powheg-pythia8/RunIISummer20UL16MiniAODv2-106X\_mcRun2\_asymptotic\_v17-v2/MINIAODSIM & 0.27 \\

    \end{tabular}
    }
    \caption{A list of the datasets used to construct AOJ and the Monte Carlo samples used to estimate its composition.}
    \label{tab:samples}
\end{table*}

\section*{Acknowledgements}
This work was initiated at the Aspen Center for Physics, supported by National Science Foundation grant PHY-2210452. We would like to thank Alexander M\"uck for discussions. The research of MK and HR-G is supported by the Deutsche Forschungsgemeinschaft DFG under grant 396021762 -- TRR 257: Particle physics phenomenology after the Higgs discovery. JB, AH and GK are supported by the DFG under the German Excellence Initiative -- EXC 2121  Quantum Universe – 390833306, and by PUNCH4NFDI – project number 460248186. DAF, IP, and DS are supported by DOE grant DOE-SC0010008. OA is supported by Fermi Research Alliance, LLC under Contract No. DE-AC02-07CH11359 with the U.S. Department of Energy, Office of Science, Office of High Energy Physics. LA is supported by the University of Bologna. Additionally, we acknowledge support from the Maxwell computational resources at Deutsches Elektronen-Synchrotron DESY, Hamburg, Germany, and computing resources provided by RWTH Aachen University under project rwth0934. This research used resources of the National Energy Research Scientific Computing Center, a DOE Office of Science User Facility supported by the Office of Science of the U.S. Department of Energy under Contract No. DE-AC02-05CH11231 using NERSC award HEP-ERCAP0027491.

\section*{Code}
The code that was used to create \textsc{AspenOpenJets} from CMS open data can be found
at \href{https://github.com/OzAmram/AOJProcessing}{https://github.com/OzAmram/AOJProcessing},
and the code used for the \oja model and its training can be found at
\href{https://github.com/uhh-pd-ml/omnijet_alpha}{https://github.com/uhh-pd-ml/omnijet\_alpha}.

\section*{Dataset}
The \textsc{AspenOpenJets} dataset can be found at \href{http://doi.org/10.25592/uhhfdm.16505}{http://doi.org/10.25592/uhhfdm.16505}.

\appendix

\section{CMS Samples}\label{app:mc}
MC samples provided by CMS were used to estimate the composition of the AOJ sample. 
CMS uses a variety of generators for the simulation different physics processes, including \textsc{MADGRAPH5\_aMC@NLO}~\cite{Alwall:2014hca}, \textsc{PYTHIA}~\cite{Sjostrand:2014zea}, and \textsc{POWHEG}~\cite{Nason:2004rx,Frixione:2007nw,Alioli:2010xd}.
Each sample uses \textsc{PYTHIA} with the underlying event tune CP5~\cite{CMS:2019csb} to simulate the parton shower and hadronization.
All samples use the NNLO NNPDF 3.1 parton distribution functions \cite{Ball:2010de,NNPDF:2014otw,NNPDF:2017mvq}. 
Some physics processes are split into several samples covering different kinematic regions. 
Each sample is normalized to the appropriate cross section. 

To estimate the number of fully merged W, Z, top, and H$\to \mathrm{b\bar{b}}$ jets the corresponding MC samples were used.
A matching criteria was applied based on truth-level information from the generator to ensure the estimate reflected the number of fully-merged jets of each type. 
For each heavy resonance, generator-level information was used to check that all the quarks from the heavy resonance decay were within $\Delta R < 0.8$ of the reconstructed AK8 jet found by our selection. 
The remaining 99.5\% of jets not estimated to be heavy resonance decays are assumed to be QCD.
The resulting numbers should be considered rough estimates, as the MC modeling of these signals is imperfect, and the estimates do not incorporate corrections to the simulated trigger and reconstruction efficiency typically employed in physics analyses by CMS. 

Table \ref{tab:samples} lists the CMS data and MC samples used and their corresponding cross sections.

\bibliography{HEPML, other}
\bibliographystyle{apsrev4-1}

\end{document}